\documentclass{aa}

\usepackage{graphics}

\begin{document}

   \thesaurus{1         
              (
               )} 
   \title{ HI and OH absorption at $z=0.89$ }
   \author{ Jayaram N. Chengalur \inst{1}
     \and   A. G. de Bruyn \inst{2,3}
     \and   D. Narasimha \inst{4}
          }

   \offprints{}

   \institute{NCRA-TIFR, PO Bag 3, Ganeshkhind, Pune 411007, India.
        \and Netherlands Foundation for Research in Astronomy, P.O. Box 2,
              7990 AA Dwingeloo,
             The Netherlands
    \and Kapteyn Astronomical Institute, P.O. Box 800, 9700 AA,
        Groningen, The Netherlands
    \and     Tata Institute of Fundamental Research, Homi Bhaba Road,
             Mumbai 400 005, India.
             }

   \date{Received 19 November 1998; accepted 11 January 1999}

   \maketitle

   \begin{abstract}

        We report on WSRT observations of HI and OH absorption at a redshift 
of $z=0.885$ towards the radio lens PKS~1830-21. The mm wave transitions of 
several molecular species have already been observed in absorption towards
PKS~1830-21 at this redshift. At mm wavelengths the source structure
is dominated by two extremely compact components, the northeast (NE) and 
southwest (SW) components. At lower frequencies the continuum emission
is much more extended and there is also a broad Einstein ring connecting the
NE and SW components. This larger extent of the continuum means that the
HI and OH spectra sample a much larger region of the absorber than the
mm wave spectra.

        The HI spectrum that we obtain is asymmetric, with a peak 
at $-147$~km/s with respect to the main molecular line redshift of 
$z=0.88582$. Weak mm wave molecular absorption has also been detected 
towards the NE component at this same velocity. The HI absorption, however,
covers a total velocity width of 300~km/sec, including velocities well 
to the red of the deep molecular features, and is fully resolved suggesting
that it is spatially widespread. In OH we detect both the 1667 and the 
1665~ MHz transitions, and the velocity-integrated ratio of their 
optical depths is consistent with what is expected in thermal 
equilibrium. The OH spectrum has a velocity width comparable to that 
of the HI spectrum, suggesting that it too is widespread in the absorber. 
The lack of a prominent HI peak in the spectrum at the velocity 
corresponding to the SW component, suggests that the galaxy responsible
for the absorption at $z=0.885$ has a central molecular disk many kpc
in size, and that HI may be deficient in this central region. 
        
        Our observations are sensitive to the large scale kinematics of
the absorber, and to first order the implied dynamical mass is consistent 
with the lens models of Nair et. al. (1993). Higher spatial resolution 
is however critical in order to better constrain the lensing models.

\keywords{
Cosmology: observations --
galaxies: abundances --
Radio lines: ISM --
ISM: abundances 
}

   \end{abstract}

\section{Introduction}
\label{sec:intro}

        Neutral gas at high redshifts is most easily observed through
the Lyman-$\alpha$ transition of the hydrogen atom, which with current
technology can be detected in absorption against the UV continuum of
QSOs even at column densities as low as low as $10^{13}$ atoms/cm$^{-2}$. 
The bulk (by mass) of the neutral gas however is found in the few 
very high column density systems(Rao and Briggs, 1993), where one could 
in principle expect a non trivial molecular fraction. However, quantitative
predictions of the molecular fraction are difficult to make since the 
conversion of gas from atomic to molecular form depends on a variety of
environmental factors like the UV background, the metallicity and the dust
content, all of which are poorly constrained  at high redshift. On the
observational front, despite searching a large sample, mm molecular lines
have been detected in absorption at high redshifts only from four sources
(of which two are gravitational lenses and two appear to arise from 
gas associated with the AGN itself)  (Wiklind \& Combes 1998). Here 
we discuss the case of PKS~1830-21, which is the brightest known 
radio lens.

        PKS~1830-21 was identified as a candidate gravitational lens on
the basis of its peculiar radio spectrum and morphology (Rao \& Subhramanyan
1988, Subhramanyan et. al. 1990,  Jauncey et. al. 1991). The radio structure
(see Figure~\ref{fig:ov}) consists of two compact flat spectrum components
separated by $\sim 1^{''}$ (henceforth called the northeast (NE) and 
southwest (SW) components respectively), joined by a steep spectrum ring.
At a frequency of 1.7~GHz roughly one third of the observed flux comes 
from the ring and each of two compact components. At the redshifted 
frequencies of HI (753 MHz) and OH (884 MHz) the ring is expected to be even
more dominant. The lack of simultaneous multi-frequency flux density
measurements of sufficient angular resolution (in view of the strong 
variability of 1830-21) makes a more accurate assessment of the ring
flux and the relative components fluxes at the low frequencies not
possible at present.

For long, no optical counter part has been found for 1830-21
(Djorgovski et. al. 1992), largely because of confusion arising from its 
low galactic latitude, although there is now some evidence for one (Courbin
et. al. 1998). Two independent gravitational lensing models have been 
proposed for PKS~1830-21, (Nair et. al. 1993, Kochanek \& Narayan 1992).
At the time that these models were made no redshift was available either
for the source or the lens.

         The redshift of the lens is now known to be $0.89$ from molecular 
line observations (Wiklind \& Combes 1996). The absorption spectra against
the NE and the SW image are very different (Frye et. al. 1996, 
Wiklind \& Combes 1998), ruling out the possibility that the molecules at 
$0.89$ are associated with the background quasar itself. The bulk of the 
molecular absorption occurs against the SW component, although much weaker 
absorption is also seen in some molecules against the NE component. The 
velocity separation between the absorption seen against the NE image and 
the SW image is 147 km/s. In addition to the molecules seen at $z=0.88582$,
HI absorption has also been seen towards PKS~1830-21, but at a lower reshift
of $0.19$ (Lovell et. al. 1996). The velocity width of this HI line is 
$\sim 30$ km/s and it has been interpreted as arising due to absorption 
in a dense spiral arm of a low redshift spiral galaxy. No molecular absorption
has been detected from this lower redshift system (Wiklind \& Combes 1997).

         In what follows we report on WSRT observations of the HI and OH 
absorption arising from the system at $z=0.89$. At mm wavelenghts only 
the extremely compact, flat spectrum components of the background source 
have sizeable flux. Consequently the spectra sample a region of order only
a few tens of parsecs across.  At the HI and OH  frequencies however, 
the background source is considerably more extended. These lines are thus 
more suited to probe the large scale kinematics of the absorbing system as
well as to determine the averaged physical properties on a kpc scale.

\section{ Observations and data reduction }
\label{sec:obs}

        The observations were done with the broad band UHF receivers installed
at the WSRT as part of the on going WSRT upgrade. The HI observations are
summarized in Table~\ref{tab:hi} and the OH observations in 
Table~\ref{tab:oh}

\begin{table}
  \caption[]{HI Observations}
  \label{tab:hi}
   \begin{tabular}{lcc} \hline\hline
      Date & Bandwidth & Channel Separation \\
           & MHz (km/s)     & (km/s)            \\ \hline
      03/Nov/96 & 2.5 (996)      & 31              \\
      15/Nov/96 & 5.0 (1992)     & 31              \\
      08/Jan/97 & 2.5 (996)      &  1              \\ \hline
   \end{tabular}
\end{table}

\begin{table}
  \caption[]{OH Observations}
  \label{tab:oh}
   \begin{tabular}{lcc} \hline\hline
      Date & Bandwidth & Channel Separation \\
           & MHz (km/s)$^{\dagger}$     & (km/s)$^{\dagger}$     \\ \hline
      17/Nov/96 & 5.0 (1696)      & 26.5              \\
      15/Dec/96 & 5.0 (1696)      & 26.5            \\ \hline
   \end{tabular}
   \begin{list}{}{}
   \item[$^{\dagger}$] The velocity scale is for the 1667~MHz transition.
   \end{list}
\end{table}

        The OH observations were made using the standard interferometric
mode and the data were reduced using NEWSTAR, the WSRT data reduction 
package. 1830-21 is spatially unresolved at the WSRT baselines. The data from
the two observing runs were added together (after applying the appropriate
heliocentric Doppler correction) and is shown in Figure~\ref{fig:oh}.
In addition, a lower resolution but larger total bandwidth spectrum was 
also obtained. This spectrum (which is not included here) is substantially 
the same as that shown in Figure 2. No
broader absorption features were detected.

        The high resolution HI spectrum, Figure 1{c} 
was obtained  using the WSRT as a compound 
interferometer~(CI), where the telescope was divided into two phased arrays 
and the output of these phased arrays was fed into the correlator. This mode
achieves high spectral resolution at the expense of losing spatial information.
However since PKS~1830-21 is not resolved at the WSRT, there is no loss of 
spatial information in the CI mode. The CI data was reduced using the WASP 
package (Chengalur 1996). 
The spectrum agrees well with that of Carilli et. al. 
(1997), apart from the region near $v\sim 0$, where their spectrum is badly
affected by interference.  The line is fully resolved and reaches a
peak optical depth of 5.5\%. 

The lower resolution HI spectra 
Figure 1a\&b were obtained in the standard interferometric 
mode and reduced using NEWSTAR. The observation on 15/Nov/96 used a much
larger bandwidth, however again no new broad absorption feature was
detected. As in the case for OH (but this time with better sensitivty
and a longer time baseline), there is no measureable difference between 
the spectra obtained over a period of $\sim 2$ months. The flux densities
were calibrated via reference to 3C48 for which we adopt a flux
of 25.5~Jy at 753~MHz and 22.7~Jy at 884~MHz, which are based on
the Baars et al. (1997) scale.

\begin{figure}
\resizebox{\hsize}{!}{\includegraphics{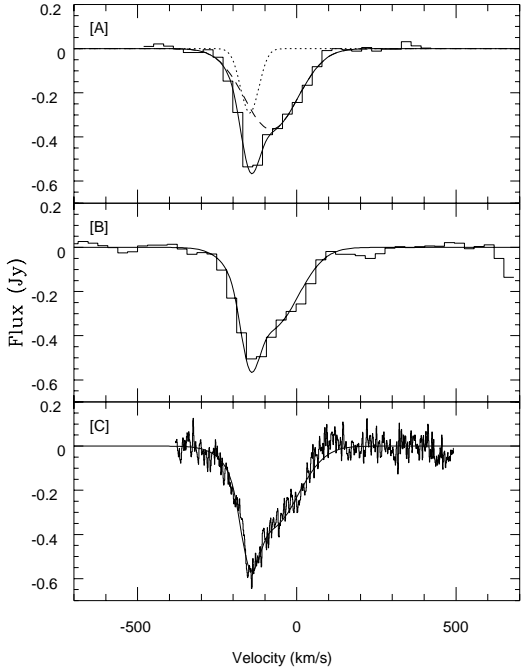}}
\caption[]{HI spectrum at $z=0.89$ towards PKS~1830-21. 
          {\bf a}~The velocity resolution is $31$~km/s. The velocity 
          scale here and throughout the paper is with respect to the 
          molecular absorption towards the SW image at $z_{hel}=0.88582$. The 
          solid line is a two gaussian fit to the spectrum, the component 
          gaussians are shown by dotted and dashed lines respectively. The 
          position and FWHM of the narrow gaussian are $-148$~km/s and 
          $40$~km/s respectively. The position and FWHM of the broader 
          gaussian is $-80$~km/s and $120$~km/s respectively. 
          {\bf b}~The HI spectrum again at a resolution of $31$~km/s, but
          with a total bandwidth of $\sim 2000$~km/s. No broad absorption 
          is seen, nor are any narrow absorption components seen at large 
          velocites. The overlaid solid line is the fit to spectrum in
          part {\bf a}. {\bf c}~The CI spectrum with a resolution of 
          $\sim 1$~km/s. Once again the superimposed solid line is not a 
          separate fit, but the same fit as in panel {\bf a}. No new narrow 
          absorption features are seen. The continuum flux of 1830-211
          at the observed frequency is about 10.5 Jy}
\label{fig:hi}
\end{figure}

\begin{figure}
\resizebox{\hsize}{!}{\includegraphics{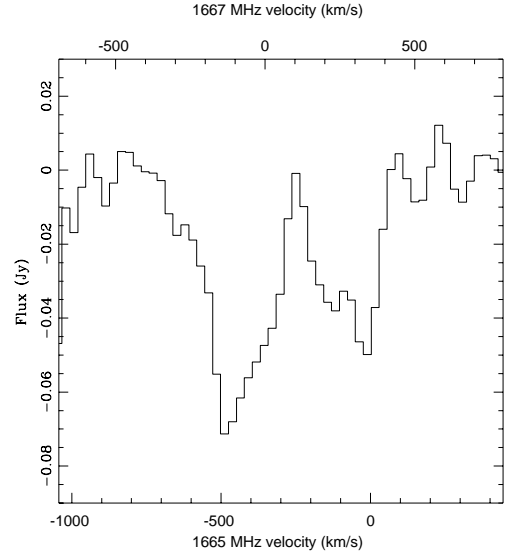}}
\caption[]{OH spectrum at $z=0.89$ towards PKS~1830-21. The lower 
           axis shows the velocity scale (with respect to $z=0.88582$) 
           for the 1667~MHz transition, while the velocity scale for 
           the 1665~MHz transition is shown in the upper axis. The 
           velocity resolution is $\sim 27$~km/s. The continuum flux
           at the observed frequency of 884 Mhz is about 10.1 Jy}
\label{fig:oh}
\end{figure}

\begin{figure}
\resizebox{\hsize}{!}{\includegraphics{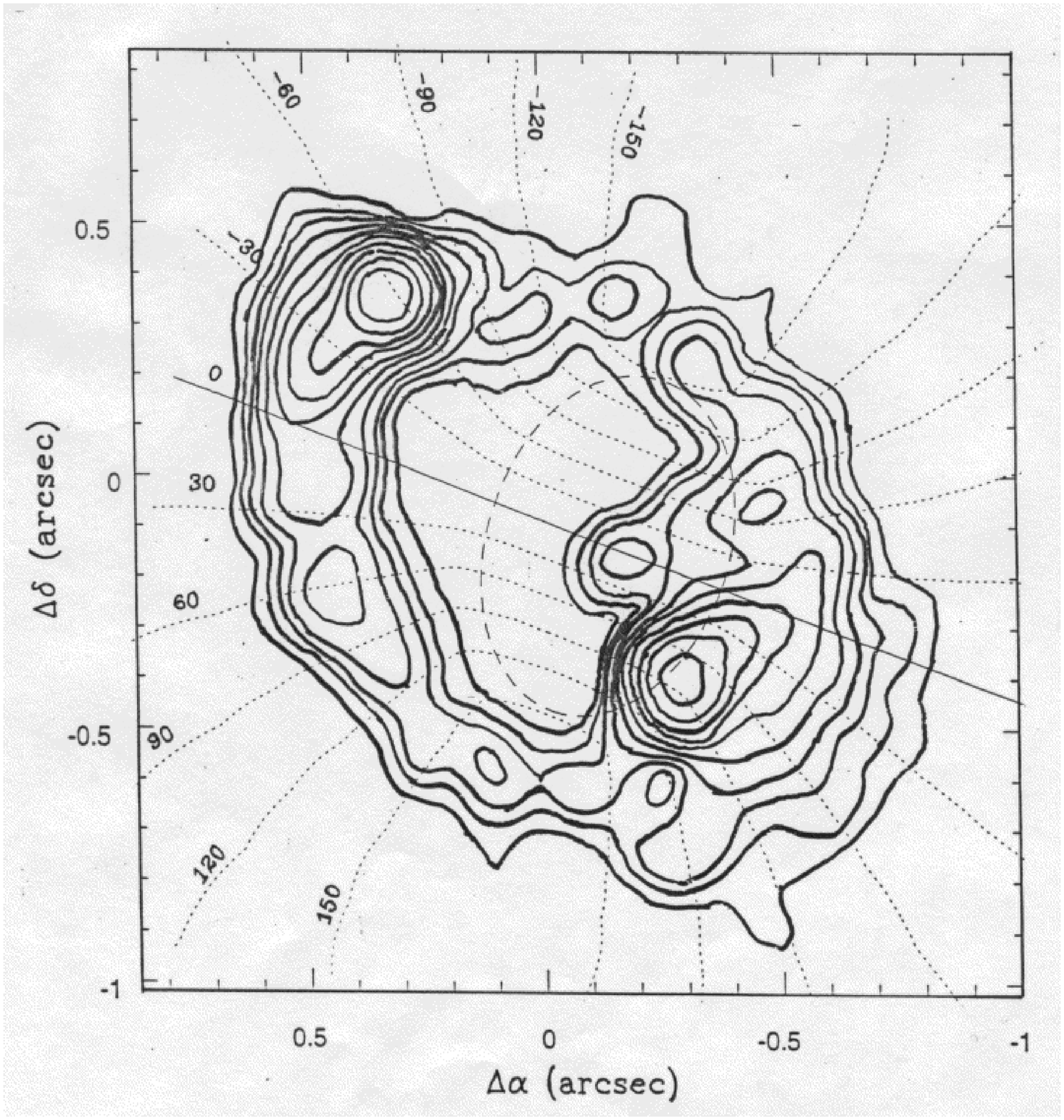}}
\caption[]{Schematic showing PKS~1830-21 superposed on a typical
           galactic velocity field. The galaxy inclination and
           position angle are close to that of the best fit model in
           Nair et. al. 1993. The radio contours are from the 6cm
	   Merlin image of Patnaik et al. 1994. }
\label{fig:ov}
\end{figure}

\section{ Discussion }
\label{sec:dis}

With peak optical depths of only 0.007 and 0.005 in the two 
OH absorption lines the profile shape is not so well defined 
as that of the HI line.
However, it is clear that the OH spectrum and the HI spectrum 
have similar overall velocity widths. Since the separation of the two 
OH lines is 350 km/sec we conclude that they do not overlap,
consistent with the height of the continuum inbetween the two absorption
features.

The 1667~MHz transition has an integrated optical depth that is larger
than that of the 1665~MHz
transition. Within the measurement errors the ratio of the
optical depth is consistent with the 9:5 ratio expected in thermal
equilibrium. There is evidence that at zero velocity 
the 1665 MHz line is deeper than the 1667 MHz line, suggesting 
variations in opacity of the 1665/1667 transitions. This could be related
to the much larger molecular line optical depth at zero velocity than
at -147 km/sec. 
We hope to address this issue with future more sensitive
observations of the OH lines.

The optical depth ($\sim 10^{-2}$), the velocity width, the overall
optical depth ratio, and the ratio of the OH column density to the excitation
temperature ($N_{OH}/T_{ex} \sim 4 \times 10^{14}$) are all within the
range of OH absorption seen towards the centers of low redshift galaxies 
(Schmeltz et. al. 1986). Under the assumption that the absorption arises 
from a rotating  galaxy disk, the large OH velocity width implies that the
covering factor of the OH absorbing gas is $\sim 1$. In the Nair et. al.
model, the distances of the SW and NE images from the lens center are 
1.8~kpc and 3.8~kpc respectively  (for $H_0 = 75$~km/s/Mpc and $q_0 = 0.5$),
and hence OH would have to be widespread in the central 5--6 kpc of
the galaxy. 

        The HI spectrum is highly asymmetric, with a peak at $-148$~km/s,
the same velocity where weak molecular absorption is seen against the NE core.
The HI peak is then presumably gas seen in absorption against the very
compact NE component. If one assumes that this component has $\sim 1/3$
of the total flux then for the gas lying in front of the NE component
$N_H/T_{sys} \sim 10^{19}$, compatible with galactic numbers of 
$N_H \sim 10^{21}$~cm$^{-2}$, and $T_{sys} \sim 100$~K. 

The red wing of the HI absorption profile shows a weak but resolved 
feature at zero velocity,  corresponding to the deep molecular absorption.
The contrast in the ratio of HI optical depth at the
two velocities, compared to that of the OH molecules, is striking.
One possibility is that the gas in front of the SW
component is primarily molecular, (i.e. similar to what is seen in many 
early type spirals).
Because the size of the radio source at 753 MHz 
is estimated to be at least 200
milliarcseconds (cf Patnaik and Porcas, 1995) corresponding about 1.5
kpc, several orders of magnitude larger than at mm
wavelengths, this lack of HI must indicate a genuine lack of HI
in a substantial part of the inner galaxy.
This, in conjunction with the OH spectrum, then suggests that the $z=0.89$
system is an early type spiral with a large central molecular disk,
at least 5--6~kpc in size. The broad component in 
the HI spectrum is presumably the result of HI seen in absorption 
primarily against the steep spectrum ring. 

        Since at low frequencies the ring has no gaps and the center
of the lensing galaxy must lie inside the ring, then without recourse
to any specific lensing model it follows that the ring must cut across
both the receding and approaching sides of 
the major axis (Figure~3). 
For reasonable rotation curves the ring will cut across the major axis well
beyond the rising part of the rotation curve. In principle then, the velocity
width of the HI spectrum ($\sim 260$~ km/s) corresponds to twice the rotation 
velocity of the  galaxy (apart from an inclination correction). In practice 
however since the emission from the ring is weakest at the points where it
cuts across the major axis, the rotation velocity could be somewhat 
underestimated. From the model in Nair et. al. (1993) it is straighforward
to compute the velocity that one should see in absorption against the SW and
the NE cores. The observed velocities are indeed obtained provided one 
changes the position angle slightly. The inclination angle in the Nair et. al.
model is $\sim 40\degr$, however this gives the mass inside the central 
4~kpc as $\sim 4\times 10^{10}~M\sun$, which is somewhat low for a
source redshift $\sim 1.5-2$. As suggested by Wiklind \& Combes (1998), the
inclination angle may be closer to $\sim 20\degr$, and the true rotation 
velocity more like $\sim 300$~km/s, more typical of early type spirals.

        In summary then, the OH and HI spectra  are consistent with the
lens being an early type spiral at a redshift of $z\sim 0.89$.

  

\end{document}